\documentclass{iopjournal}

\usepackage{hyperref}
\usepackage{fix-cm}
\usepackage{cite}
\usepackage{quantikz}
\usepackage{mathtools}
\usepackage{soul}
\usepackage{xcolor}
\usepackage{amsmath}
\usepackage{amsfonts}
\usepackage{amssymb}
\usepackage{graphicx}
\usepackage{parskip}
\usepackage{esdiff}
\usepackage{microtype}
\usepackage{mathtools}

\DeclareMathOperator*{\argmax}{argmax}

\begin{document}

\articletype{Paper} 

\title{Ground- and excited-state energies extraction via Trotterization on IBM quantum computers}

\author{Fernando Espinoza-Ortiz$^{1,*}$\orcid{0009-0005-9529-2329}, Chungwei Lin$^2$\orcid{0000-0003-1510-5414} and Chih-Chun Chien$^1$\orcid{0000-0003-1161-1846}}

\affil{$^1$Department of Physics, University of California, Merced, CA 95343, USA}

\affil{$^2$Mitsubishi Electric Research Laboratories (MERL), 201 Broadway, Cambridge, MA 02139}

\affil{$^*$Author to whom any correspondence should be addressed.}

\email{fespinoza-ortiz@ucmerced.edu}

\keywords{energy estimation, quantum computation, circuit synthesis, matchgates, controlled time evolution}

\begin{abstract}
We implement the Hadamard test with Trotterized time-evolution operators on IBM quantum computers to simultaneously extract ground- and excited-state energies of the transverse field Ising model (TFIM) and transverse longitudinal field Ising model (TLFIM). The Trotterization circuits for the TFIM admit constant-depth circuits (CDCs) for arbitrary time, allowing us to locate a large number of eigen-energies above the background noise for up to six spins. Via circuit synthesis we show that the three-spin TLFIM has constant-depth structure although it does not meet the known CDC criteria. The CDCs enable the extraction of the ground and first-excited state energies of the three-site TLFIM via its dynamics. We also address complications from the noisy background and discrete Fourier transform to enhance the reliability of the extraction process and compare the results from different generations of IBM hardware to highlight the improvement.
\end{abstract}

\section{Introduction}
Quantum algorithms are expected to outperform classical algorithms in areas such as optimization~\cite{DeSantis2026QUBO, Chatterjee2025QUBO, Lee2025QUBO, Buonaiuto2023fin, Yalovetzky2024fin}, search~\cite{Grover1996Search}, and quantum simulations~\cite{Georgescu2014QuantumSimulation, Li2015QuantumSpeedUp}. The prospect of a potential quantum speedup has motivated extensive research in the field of quantum computing: from hardware development~\cite{ransford2025helios98qubittrappedionquantum, Warner2025CoherentControl}, noise modeling~\cite{Schwartzman2025ModelError, DiBartolomeo2023NoisyGates}, error correction~\cite{Roffe2019ErrorCorrection,Acharya2025SurfaceCode}, and quantum algorithm development~\cite{Dalzell2025QASurvey, Montanaro2016Overview}. Despite the rapid progress, current noisy intermediate scale quantum (NISQ) devices~\cite{Preskill2018quantumcomputingin,RevModPhys.94.015004} are limited by coherence times and errors in qubits, gates, and measurements, restricting the application of many quantum algorithms. Nevertheless, recent experimental demonstrations highlight the growing capabilities of modern quantum computers. Notably, one of the largest quantum molecular simulations was used to compute fragmentation energies of protein-ligand complexes exceeding 12,000 atoms~\cite{Merz2026Crossing12000atombarrierheterogeneous}.

Extracting eigen spectra and weights via quantum simulation and phase estimation has become especially popular~\cite{Abrams1999EigenvaluesSpeedup, Alan2005SimulationEnergy, Dutkiewicz2022heisenberglimited, OBrien2019QpeEnergy}. Resource estimates for electronic calculation of chemical systems have been proposed for quantum phase estimation (QPE) algorithms with Qubitization~\cite{Georges2025ChemistryQubitization, Babbush2018QubitizationSpectra, Trenev2025QubitizationResource} and Trotterization~\cite{ku2025benchmarkingquantumsimulationmethods}. Both Trotterization and Qubitization~\cite{Low2019hamiltonian} have been identified as invaluable algorithms for intractable classical simulations of quantum materials~\cite{Camps2025RoadmapNercsc}. The QPE algorithm using Qubitization relies on the quantum Fourier transform~\cite{Weinstein2001QFT} to extract the energies with the precision determined by the number of ancilla qubits, controlled unitary, and multi-qubit measurements. In contrast, the Trotterization algorithm deploys the Hadamard test (H-test)~\cite{Abhijith2022QA, Mitarai2019Methodology, Cleve1998QAR} to obtain time-series data for energy extraction via classical Fourier transform. The Trotterization requires a single ancilla qubit, single-qubit measurement, and classical signal processing techniques to measure expectation values. The lower complexity of the Trotter-based algorithm and maturity of classical signal processing makes it an attractive option for extracting energy spectra on near-term NISQ devices. 

The principal challenge in implementing the H-test with Trotterized time evolution on current NISQ devices arises from the deep circuits generated by Trotterization. The first experimental realization of time-series energy extraction using the H-test with Trotterized evolution on quantum hardware was reported in Ref.~\cite{Blunt2023StatsicalPhase}, where they extract the electronic spectrum of methanethiol, $H^{+}_{3}$, $H^{-}_{3}$, and $H_{2}$. In Refs.~\cite{McKeever2023OptimizedMPO, Maurits2023CompressionMPO}, the Trotterized circuits were approximated using a parametrized circuit ansatz that are optimized through matrix product operators (MPO)~\cite{Pirvu2010MPO}. Importantly, the Trotterization circuits are compressed to constant depth with manageable gates counts. A similar MPO-based compression strategy was employed in Ref.~\cite{Kanno2026Tensor} for a variant of time-series energy estimation on a one-dimensional (1D) Hubbard model. Both studies demonstrated successful extraction of eigenenergies using constant-depth circuits (CDCs).

Motivated by the application of CDCs, the few implementations of the H-test for energy extraction, and recent findings that a subclass of Hamiltonians admits analytical constant-depth structures~\cite{BassmanOftelie2022}, we investigate the extraction of eigen-energy spectra for the transverse-field Ising model (TFIM) and the transverse-longitudinal-field Ising model (TLFIM) on two generations of IBM quantum computers using the H-test with Trotterized time evolution. The TFIM serves as a paradigmatic model for the study of quantum phase transitions~\cite{Pierre19701DTfim,Dutta_Aeppli_Chakrabarti_Divakaran_Rosenbaum_Sen_2015} and emergent thermalization phenomena~\cite{Haghshenas2026DigitalTFIM}, while the TLFIM exhibits similarly rich critical behavior~\cite{Bonfim2019GroundState, Xiao2024Discord, Lasko2021Mixed} and has also demonstrated signatures of glassy spin dynamics~\cite{Mohtashim2025TLFIMQC}, phenomena present in biological systems~\cite{Sadhukhan2021BioTLFIM}. Importantly, the TFIM belongs to the subclass of interacting spin models that admit CDCs~\cite{BassmanOftelie2022}, whereas the TLFIM does not possess a known analytical constant-depth structure. Nevertheless, we will show that the three-spin TLFIM admits CDCs through circuit synthesis using the Berkeley Quantum Synthesis Toolkit (BQSkit)~\cite{Younis2021BQSKit}.

Our approach differs from previous studies in several respects. First, the CDCs for the TFIM are directly accessible and do not rely on an ansatz. Second, the CDCs for the TLFIM are obtained through circuit synthesis, formulated as a tree-search optimization over candidate parametrized circuit structures~\cite{Smith2023LEAP}. Finally, our analysis of the resulting time signal is performed using the discrete Fourier transform (DFT). Although statistical phase estimation methods~\cite{Wan2022SPE, Somma2019SPE} are robust against noise in energy-spectrum extraction~\cite{Blunt2023StatsicalPhase}, these approaches require statistical sampling of the time signal. Given our limited quantum computing time, the DFT approach provides a more experimentally feasible alternative. Nonetheless, we simultaneously extract nine eigenenergies for the five-spin TFIM and six eigenenergies for the six-spin TFIM, surpassing previous variational approaches that reported four lowest-lying energies of a three-spin TFIM~\cite{Zhang2026LowLying}.  Our implementation extracts a large number of eigenenergies simultaneously, on par with or surpassing available results of the TFIM on accessible NISQ hardware.
On the other hand, we show that the three-spin TLFIM actually has a CDC, thereby allowing us to extract both the ground-state and first-excited state energies via its dynamics. Although eigenenergies have previously been extracted for the TLFIM of up to eight spins~\cite{Mohtashim2025TLFIMQC,hung2025improvedisingmesonspectroscopy}, those methods rely on the measurement of the magnetization and a particular relation between the energy and magnetization while the Trotterization-based method presented here is a general framework.

The rest of the paper is structured as follows. In Sec.~\ref{sec:H-test}, we review the method for simulating quantum dynamics via Trotterization, following Ref.~\cite{gunther2025phaseestimationpartiallyrandomized}. In Sec.~\ref{sec:TrotterTFIM}, we provide a gate estimate for the controlled Trotter gates of the TFIM. Sec.~\ref{sec:CDC} introduces the optimization of the CDCs for the TFIM. Sec.~\ref{sec:TrotterTLFIM}, presents the Trotterization of the TLFIM and the gate estimate for the corresponding controlled operation. In Sec.~\ref{sec:FFT}, we discuss the extraction of energy spectrum from the Fourier transform. Sec.~\ref{sec:Results} presents the results from IBM Fez and Boston quantum processing units (QPUs), first applying the H-test method to a five-spin and six-spin TFIM using the analytical CDC structure, followed by a three-spin TLFIM with global BQSkit optimization. In Sec.~\ref{sec:Discuss}, we compare our results with previous studies and discuss potential directions for future research. Finally, Sec.~\ref{sec:Conclusion} summarizes our findings and concludes our work. The Appendix summarizes some details of the CDCs and classical simulations for checking the dynamics.

\section{Background}

\begin{figure}[t]
    \centering
    \includegraphics[width=0.7\linewidth]{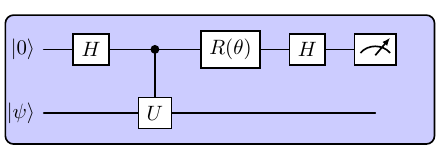}
    \caption{The Hadamard test for computing the real and imaginary parts of the time signal $g(t)$ where $R(\theta)$ is the phase gate.}
    \label{fig:htest}
\end{figure}

\subsection{Eigen spectrum estimation via Trotterization and Hadamard test}
\label{sec:H-test}
The H-test is known to extract the expectation values of a unitary operator~\cite{Abhijith2022QA, Mitarai2019Methodology, Cleve1998QAR}. The eigenenergies of any time-independent Hamiltonian $\hat{H}_0$ can be estimated with the H-test by setting the unitary operator to the time evolution operator $\hat{U}(t)$= $e^{-i\hat{H}_0t/\hbar}$. We will set $\hbar=1$ in the following. To target specific eigenvalues one may pass a \textit{guiding state} $\ket{\psi}$ to the H-test. Here we briefly review the H-test by following Ref.~\cite{gunther2025phaseestimationpartiallyrandomized}. As illustrated in Fig~\ref{fig:htest}, an ancilla qubit controls the unitary on the guiding state, and  only a measurement on the ancilla qubit is needed. The probability that the ancilla state in $\ket{0}$ is $P(\ket{0};\theta) = \frac{1}{2}[1 + \Re(e^{i\theta}\langle{\psi|\hat{U}|\psi\rangle})]$ and that in $\ket{1}$ is $P(\ket{1};\theta) = \frac{1}{2}[1 - \Re(e^{i\theta}\langle{\psi|\hat{U}|\psi\rangle})]$. 
Following the Qiskit convention, we assign the value 1 to $\ket{0}$ and -1 to $\ket{1}$, so the expectation value of the H-test circuit $\textbf{Z}_{\theta}$ is 
\begin{equation}
    \mathbb{E}\textbf{Z}_{\theta} = \Re[e^{i\theta}\langle\psi|\hat{U}(t)|\psi\rangle].
    \label{eq:qpe}
\end{equation}
Summing the outcomes of the two circuits at $\theta = 0$ and $\theta = \pi/2$ returns the complete expectation value $\langle\psi|\hat{U} |\psi\rangle = \mathbb{E}\textbf{Z}_{0}+ i\mathbb{E}\textbf{Z}_{\frac{\pi}{2}}$.

In particular, if the guiding state is represented in the eigen-basis of the Hamiltonian as $\ket{\psi} = \sum_{k}\ket{\psi_{k}}\langle \psi_{k}|\psi\rangle$, then the output of the H-test is a time-dependent function of the form
\begin{equation}
    g(t) =\langle\psi|e^{-i\hat{H}t}|\psi\rangle= \sum_{k}|c_{k}|^{2}e^{-iE_{k}t},
    \label{eq:signal}
\end{equation}
where $|c_{k}|^{2} = |\langle\psi_{k}|\psi\rangle|^{2}$ is the overlap between the guiding state and the $k$-th eigenstate of the Hamiltonian. The eigenenergies $E_{k}$ can be estimated using the Fourier transform provided that the final evolution time $\tau$ is long enough to capture the periodicity of the signal.

\subsection{Trotterization of the transverse field Ising model}
\label{sec:TrotterTFIM}
Here we study a one-dimensional (1D) $N$-spin transverse field Ising model (TFIM) with open boundary condition. The dynamics of the 1D TFIM is governed by the Hamiltonian
\begin{equation}
    \hat{H} = -J_{z}\sum_{j}^{N - 1}\sigma_{z}^{j}\sigma_{z}^{j + 1} - h_{x}\sum_{j}^{N}\sigma_{x}^{j},
    \label{eq:tfim}
\end{equation}
where $\sigma_{z}$ and $\sigma_{x}$ are the Pauli matrices, and the interactions are between nearest-neighbors (n.n.) with coupling constant $J_{z}$ and a constant transverse field along the x-direction $h_{x}$. We will use $J_z$ and $t_{0}=\hbar/J_z$ as the energy and time units in the following. 
When the Hamiltonian is a sum of Pauli operators, the time-evolution operator $U(t) = e^{-i\hat{H}t}$ can be converted into single- and two-qubit quantum gates by Trotterization~\cite{Smith2019,Lee2023Trotter}.

Since the Hamiltonian of interest is a linear combination $\hat{H}=A+B$ of non-commuting operators $[A,B]\ne0$, the decomposition of the time-evolution operator follows the Baker-Hausdorff formula~\cite{MQM}.
A naive decomposition of the exponential as $e^{-iH\Delta t} \approx e^{-iA\Delta t}e^{-iB\Delta t}$ will have a leading error of the order of $\mathcal{O}(\Delta t^{2})$, which is known as the first-order Trotterization. A symmetrization of the first-order formula ensures a leading error of $\mathcal{O}(\Delta t^{3})$~\cite{Hatano2005}. Explicitly,
\begin{equation}
    e^{-iA\Delta t}e^{-iB\Delta t} \approx e^{\frac{-i}{2}A\Delta t}e^{-iB\Delta t}e^{\frac{-i}{2}A\Delta t}.
    \label{eq:symmetrization}
\end{equation}
Although decreasing $\Delta t$ reduces the error in each step, it increases the total number of Trotter steps $M$, resulting in deeper circuits and more errors on NISQ hardware.
Therefore, $\Delta t$ 
should be chosen carefully to balance the Trotter error with the circuit depth.
If the quantum algorithm allows a CDC as explained in the next subsection, the total Trotter step and $\Delta t$ can be chosen to efficiently suppress the Trotter error.  
As explained in~\ref{App:Dt}, we present the results with proper time increments
to allows both faithful quantum dynamics and manageable quantum circuits on IBM quantum computers. 

\begin{figure}[t]
    \centering
    \includegraphics[width = 0.2\linewidth]{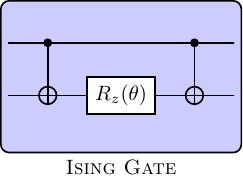}
    \caption{The $R_{zz}$ gate written with CNOT and a $R_x$ gates.
    }
    \label{fig:rzz}
\end{figure}

To investigate the scaling and qubit gate decomposition for the TFIM, we consider the first-order Trotter formula of the Hamiltonian in Eq.~\eqref{eq:tfim}. The parameters $h_{x}/ J_{z}=1$ will be used in the following. Although the parameters correspond to the quantum critical point in the thermodynamic limit~\cite{Pierre19701DTfim, Dutta_Aeppli_Chakrabarti_Divakaran_Rosenbaum_Sen_2015}, there is no singular behavior for the relatively small system studied here on quantum computers. The Trotter decomposition of the time-evolution operator for one step is
\begin{equation}
    U_{\text{tfim}}(\Delta t)\approx \prod_{j}^{N-1} e^{i\sigma_{z}^{j}\sigma_{z}^{j +1}\Delta t}\prod_{j}^{N}e^{i\sigma_{x}^{j}\Delta t}.
    \label{eq:first_trotter_tfim}
\end{equation}
The two products are combinations of two-qubit and single-qubit gates. Explicitly, there are $N$ $R_x$ gates acting on $N$ qubits: $e^{i\sigma_{x}^{j}dt} = R_{x}^{j}(\theta)$, where $\theta = -2\Delta t$. Likewise, the nearest-neighbor coupling term can be implemented by the two qubit $R_{zz}$ gate on adjacent qubits: $e^{-i\frac{\theta}{2}\sigma_{z}^{j}\sigma_{z}^{j+1}}=R_{zz}^{j, j+1}(\theta)$ with $\theta =-2\Delta t$. The $R_{zz}$ gate can be decomposed into a single $R_{z}(\theta)$ rotation sandwiched between two CNOT gates, as shown in Fig.~\ref{fig:rzz}. This decomposition must be taken into account when estimating the gate cost of the H-test, as the controlled $R_{zz}$ gate is not a native operation on IBM quantum devices even though native $R_{zz}$ is supported on newer IBM QPUs. Consequently, the transpilation process defaults to the decomposition shown in Fig.~\ref{fig:rzz} when mapping the full H-test circuit to the native gate set. A single Trotter step for the first order Trotterization of the TFIM has a total of $2(N-1)$ CNOTs and $2N-1$ single-qubit rotations, which scale linearly with the Trotter steps according to $2M(N-1)$ and $M(2N-1)$, respectively. The linear dependence is a general characteristic of Trotterization because the gates are applied in sequence.

\begin{figure}[t]
    \centering
    \includegraphics[width = 0.7\linewidth]{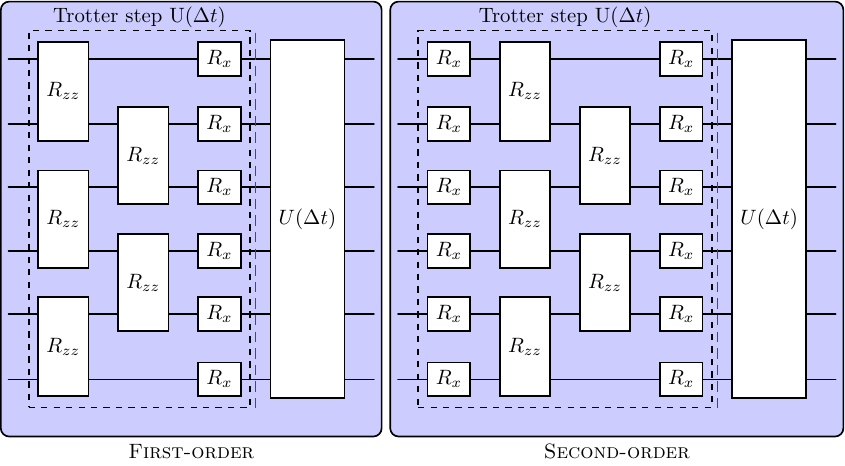}
    \caption{
    First and second order Trotterization circuits for a 6-spin TFIM with open boundary condition. Note that the rotation angle for the $R_{x}$ gates are not the same between circuits. The $R_x$ gates in the second-order Trotterization are $\theta=-\Delta t$ due to the symmetrization in Eq.~\eqref{eq:second_trotter_tfim}, and the angle in the first-order Trotterization is $\theta=-2\Delta t$. Meanwhile, the $R_{zz}$ gates share the same angle $\theta=-2\Delta t$ following equations Eq.~\eqref{eq:first_trotter_tfim} and Eq~\eqref{eq:second_trotter_tfim}.}
    \label{fig:trotterization_tfim}
\end{figure}

A solution for fewer layers of the circuit is to consider a higher order approximation with larger $\Delta t$. Although a single Trotter step may have more gates, the larger $\Delta t$ will require less Trotter steps to attain the target simulation. For the second-order Trotterization of the TFIM, a single-step time evolution becomes
\begin{equation}
    U_{\text{tfim}}(\Delta t)\approx \prod_{j}^{N}e^{i\sigma_{x}^{j}\frac{\Delta t}{2}}\prod_{j}^{N-1} e^{i\sigma_{z}^{j}\sigma_{z}^{j +1}\Delta t}\prod_{j}^{N}e^{i\sigma_{x}^{j}\frac{\Delta t}{2}}
    \label{eq:second_trotter_tfim},
\end{equation}
which introduces only a single layer of $R_x$ gates because the intermediate $R_x$s layer can be merged into a single layer. The first-order Trotterization and second-order Trotterization have the same number of CNOT gates but the second-order has one additional layer of $R_x$ gates that equates to $M(3N-1)$ single-qubit rotation. Due to the controlled unitary for the H-test, there are $2M(N-1)$ CCNOT gates and $M(3N-1)$ controlled single-qubit gates for the second-order approximations. It is known that a controlled single qubit rotation can be expressed by 3 single qubit rotations and 2 CNOT gates~\cite{Barenco1995Elementary}, and the CCNOT gate by 8 single qubit rotations and 6 CNOT gates~\cite{DiVincenzo1998Gates}. Henceforth, any element in $SU(2)$ is called a single qubit rotation and is assumed to be a native single qubit gate.

The decomposition of the controlled gates gives the total gate estimate. The number of gates for the controlled operations in the first-order approximation are approximately $22MN$ single-qubit rotations and $16MN$ CNOT gates. Our application will require a maximum of $11000N$ single qubit gates and $8000N$ CNOTs for the first-order approximation assuming $M=500$. If instead we consider the second-order scheme with twice the step size ($M=250, \Delta t/t_{0}=0.2$), it requires $6250N$ single qubit gates and $4500N$ CNOTs. Both schemes are well above the gate budget of modern NISQ devices even for $N=3$. 
Nevertheless, Ref.~\cite{BassmanOftelie2022} has shown that for certain 1D Hamiltonians, there exist
CDCs independent of the total Trotter step $M$ for the dynamics. The tradeoff is that one must search for the parameters of the fixed CDC, as explained in the following.

\subsection{Optimization to constant depth circuits}
\label{sec:CDC}
The optimization to CDCs applies only to a subclass of Hamiltonians that are quadratic in fermionic operators~\cite{BassmanOftelie2022}, for which the TFIM is included. In particular, the time-evolution operator $e^{-iHt}$ of these Hamiltonians is expressible by a quantum circuit of n.n. gates $G(A,B)$~\cite{Terhal2002matchgate}, so called matchgates. For instance, the matchgate circuit for a $M^{th}$ Trotterized circuit consisting of 4 qubits is given by a set of 4 layers of two-qubit matchgates acting on n.n. qubits~\cite{BassmanOftelie2022} 
\begin{center}
$U(\Delta t)^{M}$ = 
\begin{quantikz}[row sep=0.1cm]
&\gate[2]{G(\theta_1)}& \slice{$1^{st}$ step}& \ \ldots\ &\gate[2]{G(\theta_1)}& \slice{$M^{th}$ step}& \\
& &\gate[2]{G(\theta_2)}& \ \ldots\ & &\gate[2]{G(\theta_2)}&\\
&\gate[2]{G(\theta_3)}& & \ \ldots\ &\gate[2]{G(\theta_3)}& &\\
& & & \ \ldots\ & & &
\end{quantikz},%
\end{center}
where $\theta_i$ are their respective parameters. Henceforth, we drop $\theta_i$ but note that each $G$ corresponds to a unique parameter. 
The group structure and mirroring identity of the matchgates allow their $M$-step Trotterized circuits to be down-folded to $N$ qubit circuit with $N$ layers of matchgates (see~\ref{App:Matchgate} or Ref.~\cite{BassmanOftelie2022}). As an example, the CDC of the $M^{th}$ Trotterized circuit for a 4-qubit TFIM can be expressed as~\cite{BassmanOftelie2022}
\begin{center}
$U(\Delta t)^{M}$ =
\begin{quantikz}[slice all, row sep = 0.1cm]
&\gate[2]{G} & &\gate[2]{G}& &\\
& & \gate[2]{G}&&\gate[2]{G}& \\
&\gate[2]{G}& &\gate[2]{G}& &\\
& & & & &
\end{quantikz}.
\end{center}

However, each matchgate $G$ in the CDCs is a parametrized quantum gate with unique parameters. 
\begin{center}
\resizebox{\columnwidth}{!}{%
\begin{quantikz}
& \gate[2]{G}&\\
& &
\end{quantikz}
= 
\begin{quantikz}[row sep=0.1cm]
&\gate[1]{R_x(\theta_0)}& \gate[1]{R_z(\frac{\pi}{2})}& \ctrl{1}& \gate[1]{R_x(\theta_{1})}& \ctrl{1}&\gate[1]{R_z(-\frac{\pi}{2})}&\gate[1]{R_x(\theta_3)}& \\
&\gate[1]{R_x(\theta_0)}& \gate[1]{R_z(\frac{\pi}{2})}& \targ{}&\gate[1]{R_z(\theta_{2})}& \targ{}&\gate[1]{R_z(-\frac{\pi}{2})}& \gate[1]{R_x(\theta_3)}&
\end{quantikz}%
}
\end{center}
This introduces a set of parameters $\theta_j\in\alpha$ that must be found by numerical optimization. The objective is to search for the set of parameters $\alpha$ where the circuit unitary is approximately the unitary matrix of the time-evolution operator, derived from either direct diagonalization or Trotterization. In particular, if $C(\alpha)$ describes the unitary of the parametrized circuit and $U$ is the time-evolution operator, then the optimization problem is~\cite{Davis2020QCSynthesis, Khatri2019quantumassisted, BenDov2024Approximate}
\begin{equation}
    \argmax_{\alpha}[Tr(U^{\dagger} C(\alpha))],
    \label{eq:search}
\end{equation}
where the Hilbert-Schmidt inner product is used to define the overlap between the target unitary and the circuit’s operator. The target unitaries are chosen from the set of second-order Trotterization of the TFIM up to the final time $\tau$: $\mathcal{S} = \{ U^{2^{nd}}_{tfim}(n\Delta t) \mid n = 1, \dots, M ; M\Delta t = \tau \}$. The optimization converges when the parameters in $\alpha$ maximize the Hilbert-Schmidt inner product in Eq.~\eqref{eq:search} up to a tolerance value $\epsilon$. The optimization is solved by instantiation~\cite{Younis2022Instantiation} offered in BSQkit~\cite{Younis2021BQSKit}, where $\epsilon$ is limited to machine precision with a default value of $\epsilon=10^{-8}$.

The CDCs introduce an additional classical overhead. Mainly, for $M$ trotter steps there are $M$ corresponding optimizations. However, since each unitary in $\mathcal{S}$ corresponds to an independent CDC, they may be optimized in parallel, thereby reducing the classical overhead. Another advantage 
is that $\Delta t$ can be made infinitesimally small, provided sufficient computational resources to optimize the rapidly increasing number of CDCs. As discussed in Refs.~\cite{Matteo2016PaeallelCirc, Davis2020QCSynthesis}, the biggest challenge in optimizing the circuits is the exponential scaling with respect to the number of qubits in the circuit. Table~\ref{tab:cdc_performance} shows that after $N=8$ the optimization for a single CDC increases sharply from a few minutes to several hours. Nevertheless, for small system sizes of $N\leq8$, an order of $10^5$ CDC optimization can be generated on a 16 GB classical computer to enable a finer discretization of the time evolution, thereby reducing the Trotter error.

\begin{table}[h]
\centering
\begin{tabular}{c|c|c}
\hline
$N$ & Execution time (s) & Memory Usage (MB) \\
\hline
3 & 0.03 & 0.07 \\
4 & 0.16 & 0.14 \\
5 & 0.70 & 0.21 \\
6 & 4.48 & 0.32 \\
7 & 60.47 & 0.42 \\
8 & 1469.43 & 0.54 \\
9 & 37232.23 & 0.67 \\
\hline
\end{tabular}
\caption{Execution time and memory usage on a classical computer for the optimization of the TFIM at different numbers of spins $N$ with BQSkit's instantiation. 
}
\label{tab:cdc_performance}
\end{table}

The optimized circuits of the TFIM have $N\lfloor
\frac{N}{2}\rfloor$ number of matchgates for odd number of qubits $N$, and $\frac{N}{2}(N-1)$ matchgates for even number of qubits. They scale as $\frac{N^{2}}{2}$, for each matchgate there are 10 single qubit gates and 2 CNOTs. In the H-test, the controlled CDCs for the TFIM contain $5N^{2}$ controlled single qubit gates and $N^{2}$ CCNOTs. The decomposition of the controlled single qubit gates and CCNOTs gates into single qubit and CNOTs consists of $23N^{2}$ single qubit gates and $16N^{2}$ CNOTs. The total number of gates for the controlled CDC in the H-test of the TFIM is roughly $39N^{2}$. Although the maximum number of gates in the controlled CDC scales exponentially with the system size N, it is independent of time. In contrast, the gate count of the second-order Trotterized TFIM shown in Sec.~\ref{sec:TrotterTFIM} scales polynomially with $N$ and linearly with the evolution time. Nevertheless, for a fixed final time $\tau$, there exists a critical system size $N_{c}$ beyond which the CDC requires more gates than the corresponding Trotter circuit. For a final evolution time of $\tau/t_{0}=50$ and a time step of $\Delta t/t_{0}=0.1$ of the TFIM with the selected parameters, the critical system size is $N_{c}=282$. Since all systems considered here are much smaller than $N_{c}$, the CDCs will have less gates than the plain Trotterized circuits .
Furthermore, for $N<7$ the CDCs of the TFIM studied here will lie below the 2000 gate budget of modern NISQ devices - assuming the single qubit gates and CNOT are native gates. The scaling of the CDCs allows us to simulate the TFIM up to the required time of $\tau/t_{0}$ = 50 on IBM QPUs. However, we caution that the Hamiltonian of the TLFIM (discussed below) does not belong to the class of known CDC circuits, so more analyses and circuit syntheses are needed.

\begin{figure}[t]
    \centering
    \includegraphics[width = \linewidth]{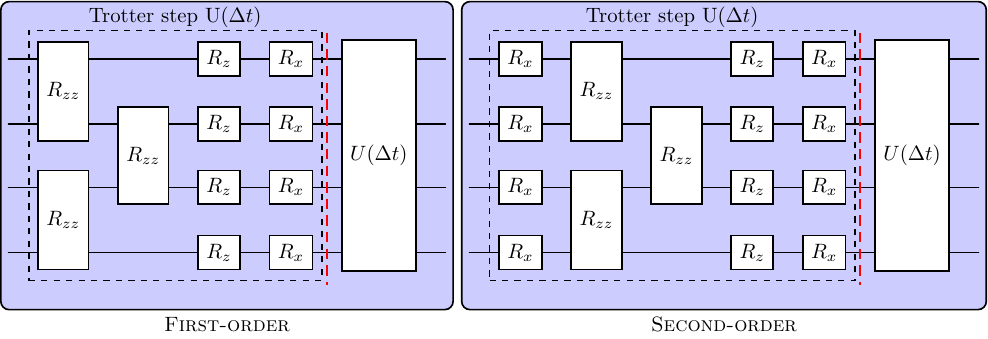}
    \caption{
    First and second-order Trotterization circuits for 4-spin TLFIM with open boundary condition. The $R_x$ gates in the second-order Trotterization have angles $\theta=-\Delta t$ due to the symmetrization in Eq.~\eqref{eq:second_trotter_tlfim}, and the angle in the first-order Trotterization is $\theta=-2\Delta t$. Akin to the TFIM, the $R_{zz}$ and $R_{z}$ gates share the same angle $\theta=-2\Delta t$ following Eq.~\eqref{eq:second_trotter_tlfim}.
    }
    \label{fig:trotterization_tlfim}
\end{figure}

\subsection{Trotterization of the transverse longitudinal field Ising model}
\label{sec:TrotterTLFIM}
The Hamiltonian of the TLFIM has an additional longitudinal field in the direction of the Ising term (z-direction in our convention):
\begin{equation}
\hat{H}_{TLFIM}=\hat{H} - h_z\sum_j^N\sigma_z^j.
\label{eq:tlfim}
\end{equation}
If $h_{z}\gg h_{x}$, the ground state of the TLFIM will align to the z-direction. Since our guiding state is the product state along the z-axis $\ket{\psi}=\ket{0}^{\otimes N}$, the ground-state contribution in Eq.~\eqref{eq:signal} will overwhelm the excited states. Therefore, we have chosen a weak longitudinal field $h_{z}/J_z=0.1$ for more spread spectral weights. The TLFIM in general is not exactly solvable and exhibits quantum phase transitions in the thermodynamic limit~\cite{Bonfim2019GroundState}. Both the TFIM and TLFIM have a second-order phase transition by the x-field, but the TLFIM has an additional first-order transition when the x-field is suppressed. The transitions happen in the thermodynamics limit, but here we analyze small systems to extract their eigen-energy spectra. We note that the TLFIM with $h_{z}/J_z=0.1$ and $h_{x}/J_z=1$ does not correspond to a critical point in the thermodynamic limit.

The second-order Trotterization yields
\begin{equation}
    U_{\text{tfim}}(\Delta t)\approx \prod_{j}^{N}e^{i\sigma_{x}^{j}\frac{\Delta t}{2}}\prod_{j}^{N-1} e^{i\sigma_{z}^{j}\sigma_{z}^{j +1}\Delta t}\prod_{j}^{N}e^{ih_{z}\sigma_{z}^{j}\Delta t}\prod_{j}^{N}e^{i\sigma_{x}^{j}\frac{\Delta t}{2}}
    \label{eq:second_trotter_tlfim}.
\end{equation}
In terms of gates, $e^{i\sigma_{x}^{j}\frac{\Delta t}{2}} = R_{x}^{j}(\theta)$, where $\theta = -\Delta t$ , $e^{-i\frac{\theta}{2}\sigma_{z}^{j}\sigma_{z}^{j+1}}=R_{ZZ}^{j, j+1}(\theta)$ with $\theta =-2\Delta t$, and $e^{i\sigma_{z}^{j}\frac{\Delta t}{2}} = R_{z}^{j}(\theta)$ with $\theta = -2\Delta t$. 
Similarly to the second-order Trotterization of the TFIM, the second-order Trotterization of the TLFIM only adds one additional layer of gates compared to its first-order approximation shown in Fig.~\ref{fig:tlfim}. Hence, we use the second-order Trotterization in the following. The number of controlled gates are $2M(N-1)$ CNOTs and $M(4N-1)$ controlled single qubit rotations, where $M$ is the number of trotter steps. This yields roughly $28MN$ single qubit rotation and $20MN$ CNOTs. The final time and time step can be chosen such that the total number of gates lie below the gate-budget of around 2000 gates. The circuit gate count is further reduced by global circuit synthesis with BQSkit ~\cite{Younis2021BQSKit}.

\subsection{Fourier spectrum and noise threshold}
\label{sec:FFT}
A continuous Fourier spectrum of Eq.~\eqref{eq:signal} ideally results in 
a sum of delta functions centered at $E_{k}$ with amplitudes $|c_{k}|^{2}$: $F_{\omega} = \sum_{k}|c_{k}|^{2}\delta(E_{k}-\omega)$, implying a real and positive Fourier spectrum. However, due to finite sampling from the quantum dynamics at discrete points, the signal in Eq.~\eqref{eq:signal} becomes $g_n = \sum_{k}|c_{k}|^{2}e^{-iE_{k}n\Delta t}$ with $n=\{ 1, \cdots\ M;\tau = M{\Delta t},M\in\mathbb{Z}^+\}$.
Additionally, the eigenenergies $E_k$ are proportional to the phase in $g_n$.
The Fourier spectrum computed by the discrete Fourier transform is not guaranteed to be delta functions. Instead, ${F}_{k'}=\sum_k |c_k|^2 \sum_{m=0}^{M-1} e^{i(E_{k} - \omega_{k'})m\Delta t}$ with $k'=0, \cdots, M-1$. Only when $E_{k}$ is a multiple of $\Delta\omega=\frac{2\pi}{M\Delta t}$, the summation returns a delta function. Therefore, the amplitudes in the discrete Fourier spectrum can be complex valued. Following Ref.~\cite{Sinha2025Lecturesquantumfieldtheory}, we present the absolute value of the Fourier coefficient, which includes the information in both real and imaginary parts of the spectrum. 
Equivalently, the sequence $g_n$ can be expressed as
\begin{equation}
    g_{n} = \frac{1}{M}\sum_{k'=0}^{M-1}F_{k'}e^{i2\pi k'n/M}.
    \label{eq:ift}
\end{equation}
The sum over the entire Fourier spectrum must conserve probability: $\sum_{k'}\frac{F_{k'}}{M}= \sum_{k}|c_{k}|^{2}=1$.  
We caution that this time-series analysis to extract the energy spectrum is limited to gapped Hamiltonians with a discrete energy spectrum.

Additionally, we found that the ratios $\frac{|F_{k'}|}{|F_{max}|}$ of the few largest-weighted eigen-energies in the spectrum normalized by the maximal-weight value in the Fourier spectrum reflect the corresponding values of $\frac{|c_{k}|^{2}}{|c_{max}|^{2}}$ found by exact diagonalization on classical computers. Thus, we plot the ratio $\frac{|F_{k'}|}{|F_{max}|}$ of the discrete Fourier spectrum to locate the eigenenergies and to retrieve relative information about the weights of the few eigenstates with significant overlaps with the initial state.


We note that the eigenenergies from a finite-dimensional Hamiltonian correspond to the peaks in the Fourier spectrum within some interval because the eigenvalues of a bounded Hermitian operator is confined to its operator norm~\cite{Demuth2015BanachBounded, Domokos2013HermitianBounded}. As a consequence, the step-size $\Delta t$ must follow the Nyquist rate ($1/\Delta t>\frac{E_{max}}{\pi}$) to ensure that the discrete signal reproduces the continuous signal without loss of information~\cite{Maurice1992Shannon, Jerri1977Shannon}. 
Moreover, the frequencies are given by $\frac{2\pi k'}{M\Delta t}$ for integer $k'$, and setting $k'=1$ gives the smallest measurable frequency $|E_{min}|$ of the spectrum. We choose the time-step $\Delta t$ to allow faithful dynamics while simultaneously satisfying the Nyquist rate. 
By varying the initial states $\ket{\psi}$ with different overlaps of the ground- and excited states of the target Hamiltonian, different eigenvalues can be found through the Fourier spectrum. 

Hitherto, we have neglected the noise from the NISQ hardware which results in fluctuations across all frequencies in the Fourier spectrum. The presence of noise results in a noisy background across the Fourier spectrum, calling for a threshold value $\eta$ to distinguish signals from background noise~\cite{Schoukens1986DFTNoise}. Since we cannot control the activity of neighboring qubits between runs on the hardware, and therefore errors associated with crosstalk~\cite{Sarovar2020Crosstalk, Alghadeer2026Crosstalk,Lange2026Crosstalk}, using a global value of $\eta$ to filter our spectrum is untenable. Rather, $\eta$ should be derived with respect to the spectrum's own background for each run on quantum computers. To this end, we define $\eta_{i}$ as the highest peak in the tails of the Fourier spectrum when the energy scale $\frac{2\pi k't_{0}}{M\Delta t}\gg \frac{E_{N,max}}{J_{z}}$ (an estimation of the maximal eigenenergy of the system of interest). The subscript $i$ emphases the noise threshold is defined with respect to the specific set of data. 
Explicitly,
\begin{equation}
    \eta_{i} = \max\{\frac{|F_{k'}|}{|F_{k'}^{max}|};E_{k'}\notin[E_{min}, E_{max}]\}
    \label{eq:background}.
\end{equation}

Given any Hamiltonian, $E_{min/max}$ can be estimated using Geršgorin bounds~\cite{Richard2010Circles, Slepian1978EigenvaluesHermitian, Deville2019Gershgorin}. Only peaks in the Fourier spectrum with amplitudes greater than the threshold $\eta_{i}$ are considered as successfully extracted eigenenergies. For the spin models studied here, we have access to the entire eigenenergy spectrum from exact diagonalization on classical computers. For a fair comparison with the analytic results, the exact eigenenergies are plotted as vertical lines in the Fourier spectrum with amplitudes normalized by that of the largest weighted eigenstate $\frac{|c_{k}|^{2}}{|c_{max}|^{2}}$. 
The conservation of probability, $\sum_{k'}\frac{F_{k'}}{M}=1$, still holds for noisy spectrum,
and $\sum_{k'}\frac{|F_{k'}|}{M} \ge 1$ because of the triangle inequality~\cite{Phillips2020Complex}.

\section{Results}
\label{sec:Results}
Here we present the Trotterization-based eigenenergy estimation of the TFIM and the TLFIM on the older IBM Fez the newer IBM Boston whenever CDCs are obtained. The guiding state for all circuits is chosen to be the product state in the computational basis, $\ket{0}^{\otimes N}$, which avoids complexity from state preparation while giving reasonable distributions of spectral weights for the ground state and first few excited states.

\begin{table}[t]
\centering
\begin{tabular}{c|c|c}
\textbf{Number of Spins}
& \textbf{Unitary}
& \textbf{Controlled Unitary} \\
\hline
3  & 36  & 448  \\
4  & 69  & 767  \\
5  & 116 & 1342 \\
6  & 156 & 1693 \\
7  & 217 & 2444 \\
\hline
\end{tabular}
\caption{The total ISA gate count for the CDCs of the TFIM when $R_{zz}$ and $R_x$ are included in the ISA. The circuits considered are the CDCs (labeled ``Unitary'') and their controlled operations (labeled ``Controlled Unitary'').}
\label{tab:tfim_isa_gate_count}
\end{table}

\subsection{Implementation and extraction for TFIM on IBM}
Before execution on IBM's hardware, the circuits must be transpiled into the instruction set architecture (ISA) of the selected QPU by converting the operators into the native gates. This step alongside qubit routing, and qubit selection is performed with Qiskit's pass manager at optimization level 2. For IBM Fez and Boston QPUs, the pass manager selects the optimal topology of n.n. qubits and has a set of native gates of CZ, $R_{x}$, $R_{z}$, $R_{zz}$, $S_{x}$, and X. We caution that by default $R_{x}$ and $R_{zz}$ are not included in the ISA of IBM Fez and Boston, and the user must opt-in to include them in the ISA. 
There are several restrictions when using fractional gates ($R_{zz},R_x$)~\cite{ibm_fractional_gates}: The main conflict is that fractional gates do not support dynamic circuits (mid-circuit measurements), Pauli twirling, probabilistic error cancellation, and zero-noise extrapolation. Here, we include the fractional gates in the ISA of IBM Boston and Fez, since no error correction or mitigation is applied; as the CDCs of the TFIM with the H-test are within the NISQ limit for both $N=5$ and $N=6$ TFIM (see Table.~\ref{tab:tfim_isa_gate_count}). Meanwhile, the angle of the $R_{zz}$ gate is constraint to $0<\theta<\frac{\pi}{2}$, which requires a custom pass manager that adds single qubit gates to the $R_{zz}$ gates if the angle does not fall in the default range by Qiskit's FoldRzzAngle.

We extract the time signal $g(t)$ of Eq.~\eqref{eq:signal} for $N=5,6$ TFIM on IBM Fez and IBM Boston using the H-test in Fig.~\ref{fig:htest}. The unitary gate is the CDC of the TFIM (Eq.~\eqref{eq:tfim}) found by solving Eq.~\eqref{eq:search}. Here we evolve each spin system to a final time of $\tau/t_{0}=50$ which corresponds to a frequency spacing of $|\frac{E_{min}}{J_{z}}|\approx0.1$ and requires 500 CDC optimization for a step-size of $\Delta t/t_{0}=0.1$. Importantly, the time increment satisfies the Nyquist rate ($t_{0}/\Delta t>\frac{E_{max}}{J_{z}\pi}$)~\cite{Maurice1992Shannon, Jerri1977Shannon} which assures that the largest eigenvalue in the Hamiltonian's spectrum can be resolved without aliasing. The energy spectra are extracted by discrete Fourier transform of Eq.~\eqref{eq:signal}, and we reiterate that the exact, continuous Fourier spectra are real-valued but the discretization causes complex coefficients. Consequently, the modulus is presented to reflect the total strength of the signals. 


\begin{figure}[t]  
  \centering
  \includegraphics[width=0.7\textwidth]{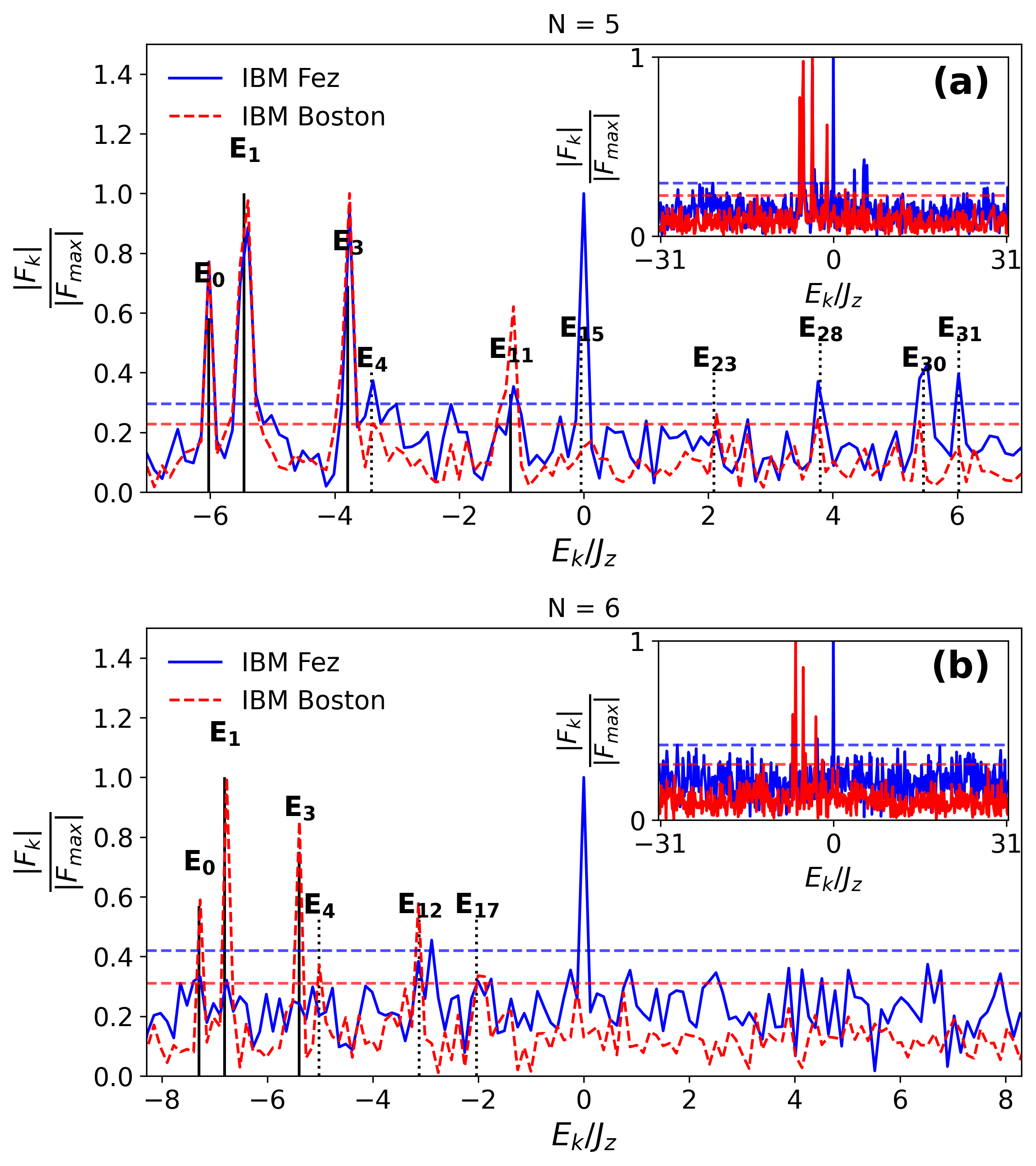}  
  \caption{Fourier spectra of the TFIM with $h_x/J_z=1$, $\Delta t/t_{0}=0.1$, and $\tau/t_{0}=50$ for (a) $N=5$ and (b) $N=6$ spins from IBM Fez and Boston QPUs. The circuits were ran with 6500 shots per circuit to ensure signal resolutions. Each spin simulation consisted of 1002 total circuits; split evenly for extracting the real and imaginary components of the signal. With the guiding state $|0\rangle^{\otimes N}$, for $N=5$ ($N=6$) we can resolve 9 (6) eigenenergies, indicated by the vertial dashed lines. The horizontal dashed lines indicate the background noise threshold according to Eq.~\eqref{eq:background}.
  }
  \label{fig:estimation}
\end{figure}

The Fourier spectra of the TFIM with $N=5,6$ are presented in Fig.~\ref{fig:estimation}. The vertical black solid lines represent eigenenergies with normalized amplitudes $|F_{k'}|/|F_{k'}^{\max}|$ exceeding the noise background threshold $\eta_i$, which align with the eigenvalues obtained from exact diagonalization on classical computers. The additional vertical black dashed lines correspond to eigenenergies from exact diagonalization whose amplitudes are supposed to be below the noise threshold yet are prominent on at least one of the IBM quantum computers. For $N=5,6$ TFIM, we are able to reliably extract the three lowest lying eigenenergies on both IBM Fez and Boston QPUs (see Fig.~\ref{fig:estimation}). Furthermore, they preserve the relative ratio with respect to the largest amplitude, showing quantative accuracy of the Trotterization-based method. 

We also notice that there are peaks above the noise background at higher eigen-energies. This result is most notable for the spectrum of $N=5$ in Fig.~\ref{fig:estimation}, where IBM Fez enhances the peaks for $E_{28}$, $E_{30}$, and $E_{31}$.
A similar phenomenon was observed in Ref.~\cite{Gu2023NoiseResilientQPE}, where the Fourier peaks of degenerate eigenvalues split due to noise, but the models studied here do not have degenerate eigenvalues. The amplification of those peaks at higher eigen-energies from NISQ computers may be a hardware-specific issue.

For the $N=6$ TFIM results shown in Fig.~\ref{fig:estimation}(b), the spectrum from IBM Fez shows a sharp peak centered at $E_{0}=0$, which originates from a correspondence between noise and the zero-frequency peak. As seen in Table~\ref{tab:tfim_isa_gate_count}, large system sizes $N$ have increasing gate counts. The deeper circuits drive the system towards a featureless, maximally mixed state described by the density matrix $\rho=\frac{1}{2^{N}}\mathbb{I}$. When the system takes this form of density matrix, there is no time-evolution due to the Liouville equation of motion $\dot{\rho}=i[\rho, H]=0$. The resulting time-series signal is then flat, masked by noise from the QPU. As a result, the Fourier transform of the noisy spectrum features a prominent peak centered at $E_0 = 0$ (as shown in the more noisy IBM Fez data in Fig.~\ref{fig:estimation}(a)), corresponding to a constant-in-time signal due to accumulated errors. In contrast, no such zero-frequency peak arises in the spectrum from the less noisy IBM Boston QPU in Fig.~\ref{fig:estimation}(b). For this reason, it is recommended to skip the zero-frequency peak as an eigenenergy and instead use it as an indicator of the quality of the QPU.

\subsection{TLFIM with circuit optimization}
We consider the parameters $h_{x}/J_z=1$ and $h_{z}/J_{z}=0.1$ in Eq.~\eqref{eq:tlfim} of a $N=3$ TLFIM and target the ground-state energy ($E_{0}/J_{z}=-3.6$) and the first-excited state ($E_{1}/J_{z}=-2.6$). To resolve the ground-state energy the time-step must fall below $\frac{\Delta t}{t_{0}} <0.88$. Our simulations were performed with a smaller step size of $\Delta t/t_{0}=0.241$ for better frequency resolution, using $M=50$ Trotter steps for a final evolution time of $\tau/t_{0}\approx 12$ and $|\frac{\Delta E_{min}}{J_{z}}| \approx 0.5$. We comment that the final evolution time and time-step are greater than those considered in the TFIM because the TLFIM are not expected to have CDC structure in the first place. Such a setup corresponds to a maximum gate count of 7200 gates for the $N=3$ TLFIM: 3000 CNOTs gates and 4200 single-qubit rotations. The Trotter based circuits of the TLFIM are well above the error budget of current NISQ devices, thus we further optimize the circuits of the TLFIM via circuit synthesis with BQSkit.

BQSkit's global optimization~\cite{Younis2021BQSKit} uses an assortment of synthesis algorithms~\cite{Smith2023LEAP, Davis2020QCSynthesis, Younis2021Qfast, Kukliansky2023Qfactor} with circuit partitioning~\cite{Liu2023PAM, Wu2021Reoptimization} to produce the best quality quantum circuits with possible lower gate counts. The default synthesis method is the LEAP algorithm~\cite{Smith2023LEAP}, which works best for circuits of up to 6 qubits. LEAP works as a search over a tree of possible parameterized circuit structures, which navigates the paths and selects the best solution. While instantiation optimized with the Hilbert-Schmidt inner product, the LEAP algorithm uses the Hilbert-Schmidt distance with the same tolerance of $\epsilon=10^{-8}$. We comment that optimization at different values of $\epsilon$ has been studied in Ref.~\cite{Mastandrea2026Intermediate}, which found the default ($\epsilon=10^{-8}$) works well. Notably, global optimization only takes BQSkit quantum circuits as inputs. BQSkit offers functions that convert popular quantum frameworks into BQSkit's, such as qiskit\_to\_bqskit, which converts Qiskit circuits into BQSKit circuits. Likewise, the optimized circuits can be converted back into the framework of choice. By default, the optimized circuit are given in terms of CNOTs and three-angle single-qubit rotation gates $U3$. However, BQSkit offers preset machine models based on real hardware and has the capability to build custom models that return the optimized circuits in the native gate set. Here, the optimizations are performed on the non-transpiled Qiskit circuits and later passed to Qiskit's pass manager for compilation.

We perform circuit optimization on the second-order Trotterization of the time-evolution operator $U$ for the TLFIM  (Fig.~\ref{fig:trotterization_tlfim}). For controlled unitaries, it is recommended to optimize the unitary gate versus the entire controlled gate because the control introduces an extra qubit and deeper circuit, which complicates the optimization~\cite{Matteo2016PaeallelCirc,Wu2021Reoptimization}. Furthermore, due to the Hilbert-Schmidt distance, the circuits produced by BQSKit are equivalent up to a global phase, which introduces a unique relative phase between the $\ket{0}$ and $\ket{1}$ states of the ancilla qubit at different time $t$: $U(t)= e^{i\phi_t}V(t)$. As a result, the time signal from the optimized circuit obtains a phase
\begin{equation}
    \tilde{g}(t) = \sum_{k}|c_{k}|^{2}e^{i\phi_{t}}e^{-iE_{k}t}, 
    \label{eq:signal_phase}
\end{equation}
where $e^{i\phi_t}$ denotes the global phase associated with the optimized Trotter circuit at the discrete time step $t = n\Delta t$. We emphasize that $\phi_t$ is not a continuous time-dependent phase; rather, each Trotter step possesses its own distinct global phase. The signal $g(t)$ is recovered by determining the global phase factor for each circuit according to $e^{i\phi_t} = \frac{V_{ij}(t)}{U_{ij}(t)}$, where $U_{ij}(t)$ is a matrix element of the time evolution operator. $U_{ij}(t)$ can be obtained directly from the second-order formula defined in Eq.~\eqref{eq:second_trotter_tlfim} or by compiling the corresponding Trotter circuit in Fig.~\ref{fig:trotterization_tlfim} to obtain the circuit's unitary. Likewise, $V_{ij}(t)$ is the corresponding matrix element of the optimized circuit unitary generated by BQSkit. The corrected signal is then obtained by removing the global phase at each time step, $g(t) = \frac{\tilde{g}(t)}{e^{i\phi_t}}$.

\begin{figure}[t]
    \centering
    \includegraphics[width=0.7\linewidth]{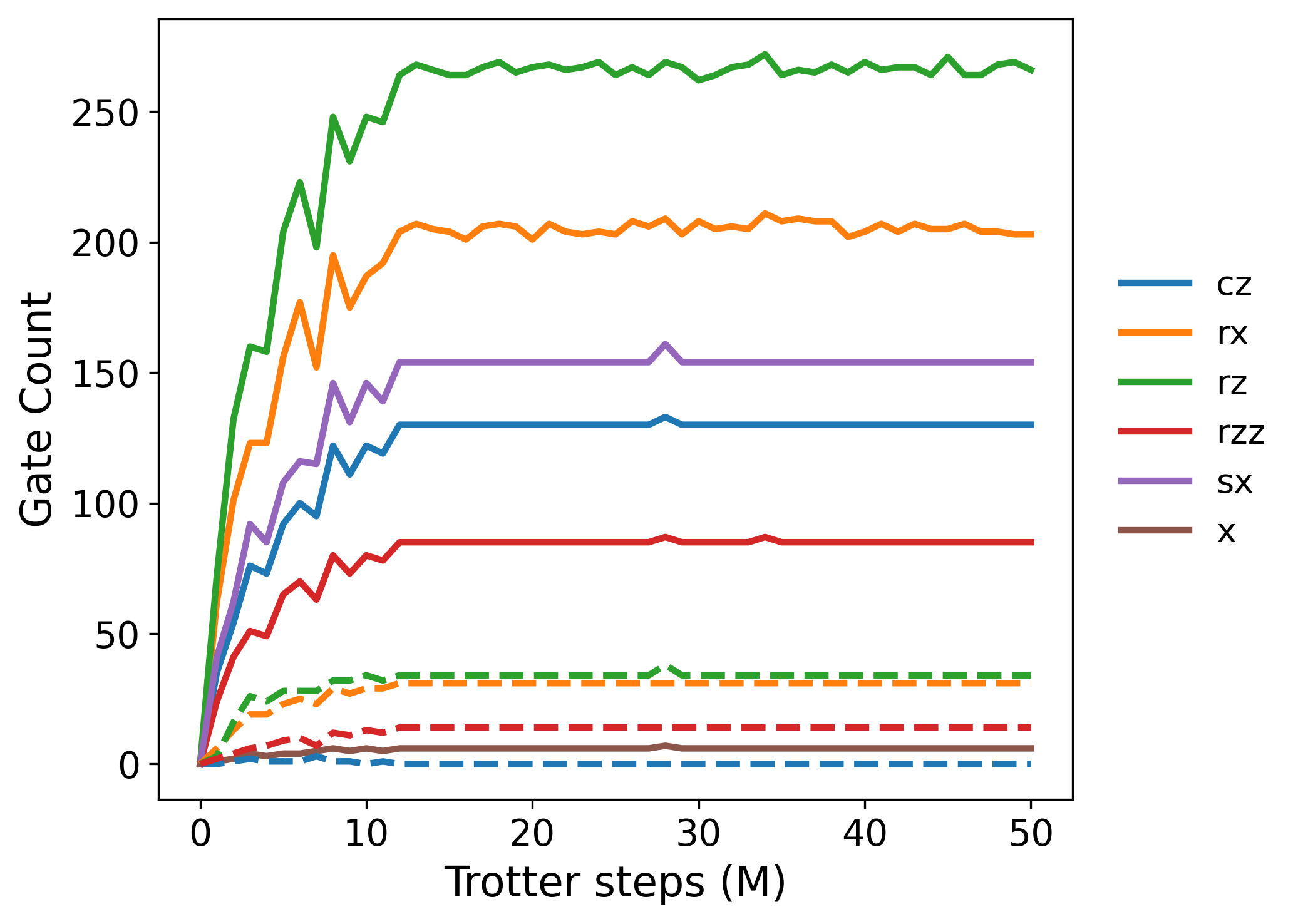}
    \caption{Gate counts versus Trotter steps $M$ for the optimized circuits of the $N=3$ TLFIM with second-order Trotterization and $\Delta t/t_0=0.241$, $h_x/J_{z} = 1$, and $h_z/J_{z}=0.1$. The dashed lines are the gate counts only for the Trotterization of the time-evolution operator after transpilation on IBM Boston, and the solid lines are the gate counts for the controlled unitary for the H-test after transpilation on IBM Boston. For the $N=3$ TLFIM, the circuit becomes of constant depth after BQSkit optimization.
    }
    \label{fig:tlfim_gates}
\end{figure}

Contrary to the typical criteria for CDCs~\cite{BassmanOftelie2022}, the Trotterized circuits of the three-spin TLFIM after the optimization exhibits constant gate counts that lie well below the NISQ limit after global BQSkit optimization; see Fig.~\ref{fig:tlfim_gates}. The emergence of CDCs in the $N=3$ TLFIM is an important finding because its Hamiltonian in Eq.~\eqref{eq:tlfim} does not map onto quadratic fermionic operators (see~\ref{App:fermion_TLFIM}), and the corresponding quantum circuits are not composed of matchgates. Consequently, the properties associated with circuit downfolding are absent in the circuits. One may suspect that the CDC arises from a small value of $\frac{h_{z}}{J_{z}}=0.1$, where the TLFIM may be close to the TFIM. However, we have verified that the CDCs persist for larger values of $\frac{h_{z}}{J_{z}}=0.5$ and $1.0$ for the $N=3$ TLFIM via BQSkit optimization as well.

\begin{table}[t]
\centering
\begin{tabular}{c|c|c}
\textbf{Number of Spins}
& \textbf{Unitary}
& \textbf{Controlled Unitary} \\
\hline
3 (CDC) & 83   & 848   \\
4       & 828  & 13123 \\
5       & 1118 & 17693 \\
6       & 1395 & 21936 \\
\hline
\end{tabular}
\caption{The total ISA gate count for the Trotterization of the TLFIM when $R_{zz}$ and $R_x$ are included in the ISA. Each circuit has been optimized with BQSkit global optimization. For $N=3$, BQSkit returns constant-depth circuits. 
We report the maximum number of gates obtained with $\Delta t/t_{0} = 0.241$ for $M=50$ Trotter steps, where the final time is $\tau/t_{0}\approx12$. The circuits considered are the Trotterized evolutions (labeled ``Unitary'') and their controlled operations (labeled ``Controlled Unitary'').}
\label{tab:tlfim_isa_gate_count}
\end{table}

In contrast, the optimized circuits for $N\ge 4$ TLFIM do not exhibit constant gate counts as far as we can check with our available resources. This may be a consequence of a larger parameter space, which results in deeper tree search. It is possible that the CDC solutions of the $N\ge 4$ TLFIM exist but the currently available optimization algorithm cannot converge to the CDC solutions because of the exponentially increasing circuits structures. A similar approach can be taken as in Ref.~\cite{Smith2023LEAP} to test the existence of a CDC for $N\ge 4$ TLFIM, where a constant-depth template from the $N=3$ TLFIM initializes a seeded search for $N=4$~\cite{Weiden2023Seeded} to hopefully improve the synthesis search. However, this is a proposal that has not been verified. Thus, optimized circuits for $N>3$ TLFIM still scale as plain Trotter-based circuits, and their possible CDCs remain elusive. Table~\ref{tab:tlfim_isa_gate_count} summarizes the gate counts for the TLFIM, and one can see that due to the CDC for $N=3$, its gate count is substantially lower and fits the error budget of NISQ hardware.

\begin{figure}[t]
    \centering
    \includegraphics[width = 0.7\linewidth]{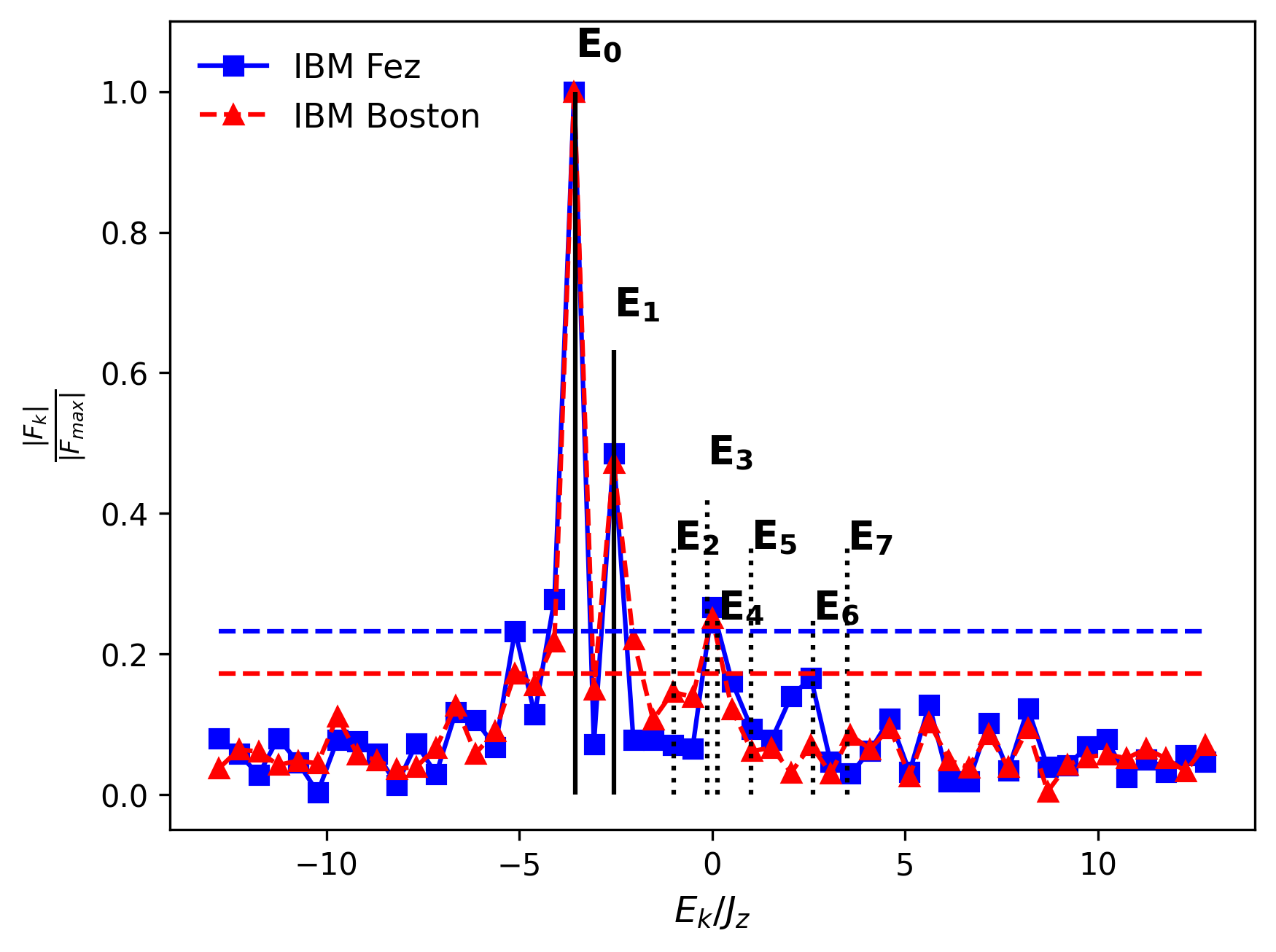}
    \caption{Fourier spectrum of the $N=3$ TFLIM using second-order Trotterization with global BQSkit optimization and $\Delta t/t_0=0.241$, $h_x/J_{z} = 1$, $h_z/J_{z}=0.1$, and $M=50$ Trotter steps. With BQSkit optimization, we can resolve the ground-state energy ($E_{0}/J_z=-3.6$) and the first-excited state energy ($E_{1}/J_z=-2.6$), indicated by the vertical solid lines. The horizontal dashed lines are the noise thresholds determined from the background.
    }
    \label{fig:tlfim}
\end{figure}

The CDCs for the $N=3$ TLFIM from BQSkit optimization enables simulations of quantum dynamics for long enough time to extract both ground-state and first-excited state energies, as shown in Fig.~\ref{fig:tlfim}. 
Since the eigenvalues $E_3$ and $E_4$ are very close to each other in this case, their difference lies within the frequency resolution. The third peak slightly above the noise threshold from the QPUs in Fig.~\ref{fig:tlfim} thus cannot unambiguously distinguish the third and fourth eigenenergies.
Consequently, only the two lowest eigenenergies (labeled by the vertical black solid lines) can be reliably extracted from the TLFIM for this setting while higher eigenenergies (labeled by the vertical dashed lines) are not resolvable. Since the $N=3$ TLFIM shows CDC scaling to improve spectral resolution substantially, the example makes the Trotterization-based approach more promising. However, finding the CDCs for $N>3$ TLFIM will be a challenging task crucial for pushing the Trotterization based method to larger size.

\section{Discussion}
\label{sec:Discuss}
Previous studies have extracted the eigenenergies of the TLFIM for systems of up to eight spins by performing Fourier analysis on the magnetization of a single spin~\cite{hung2025improvedisingmesonspectroscopy, Lamb2024}. These works report a single dominant energy in the spectrum, and in some cases a second weaker spectral peak, for various initial states. However, this approach is not meant to be a general framework but specific to the TLFIM with particular parameters and relies on the underlying conformal field theory description of stable mesons whose spectrum can be extracted from the dynamics of the magnetization. Other works have implemented variational quantum eigensolvers (VQEs) to extract eigenenergies. Ref.~\cite{Duriez2025VQETLFIM} extracted the ground-state energy of up to $N=12$ TLFIM, and Ref.~\cite{Ashutosh2026VQETFIM} solves for the ground-state of a $N=10$ TFIM. Although VQEs have found success in extracting eigenenergies, they are usually limited to only the ground or the highest-excited energy levels. Recently, a VQE that extract simultaneous eigenenergies has been implemented for a three-spin TFIM~\cite{Zhang2026LowLying}, which can extra the 4 lowest eigen-energies. Although promising, VQEs are subject to the barren plateau problem~\cite{Larocca2025Barren}, where the optimization landscape becomes featureless with increasing system size. 

Since the Trotterization-based method with the H-test is non-iterative and applicable to general systems, we are able to extract nine eigenenergies for five-spin TFIM, five eigenenergies for six-spin TFIM, and two eigenenergies for three-spin TLFIM via their quantum dynamics using the H-test with constant-depth Trotter circuits, all using the product state of the computational basis as the guiding state. The energy spectra of the TFIM not only captures multiple eigenenergies but also the ratio of their amplitudes with respect to the largest weight $|c_{max}|^{2}$. Furthermore, the framework and implementation presented here can be applied to larger and more complicated systems, assuming more computational resources are available. To the best of our knowledge, Ref.~\cite{Blunt2023StatsicalPhase} is the first implementation of energy estimation via Trotterization and the H-test extracting three eigenenergies of the molecules $H^{+}_{3}$, $H^{-}_{3}$, and $H_{2}$. Our implementation serves as an updated proof-of-concept for Trotterization-based energy extraction and demonstrates that accurate spectral information can be extracted from time-series analysis using modest circuit resources on NISQ quantum hardware. In particular, the ability to resolve multiple low-lying eigenenergies and their associated spectral weights highlights the utility of time-domain approaches for probing many-body quantum systems beyond simple ground-state estimations.

While the present study remains limited by Trotter step errors, final evolution time, and decoherence, the approach can be systematically improvable through higher-order Trotter formulas~\cite{Hatano2005, Ostmeyer2023Trotter, malezic2026efficienttrottersuzukischemeslongtime} and error mitigation techniques~\cite{Blunt2023StatsicalPhase}. We caution that the implementation of error mitigation techniques on IBM hardware requires the exclusion of the fractional gates from the ISA. Future work may consider the implementation of randomized time-evolution algorithms ~\cite{gunther2025phaseestimationpartiallyrandomized, Wan2022SPE} via CDCs, which may further reduce the number of Trotter steps for near-term applications. Similarly, the use of advanced signal processing techniques, such as compressive sensing~\cite{Needell2009TopicsCompressedSensing}, avoid the Nyquist sampling theorem and may also result in fewer Trotter steps~\cite{Yoshimura2014DiabaticRamping}. More recently, the efficient construction of the controlled time-evolution operator introduced in Ref.~\cite{Fujiwara2026EfficientContolled} has significantly reduced the gate count, making the H-test based method more competitive. 



\section{Conclusion}
\label{sec:Conclusion}
We have presented a proof-of-principle implementation of the H-test with Trotterization of quantum dynamics for energy-spectrum estimation on IBM quantum computers with exemplary spin models based on the Fourier analysis of time-series data. In particular, the Trotterization circuits are found to be of constant depth for the TFIM in general and for $N=3$ TLFIM, which enables the scheme on IBM Fez and Boston processors within the tight error budgets. We successfully extract the ground- and excited-state energies for five-spin and six-spin TFIMs and a three-spin TLFIM above the noisy background. 

Importantly, the Trotterization scheme extracts simultaneously multiple eigenenergies, showing its competitiveness compared to other available methods for utility-scale NISQ computation. With more resources available in the future, exploring the possible CDCs for larger TLFIM and applying the framework to other spin or electronic systems will further demonstrate the versatility and capabilities of the optimization and implementation of the Trotterization-based method. Moreover, our work serves a series of optimization for energy-spectrum extraction on NISQ devices to bridge time-series algorithms with early fault-tolerant quantum computation~\cite{Ding2023simultaneous} capable of processing deeper circuits.

\ack{
C.C.C. was supported by the US Department of Energy under grant No. DE-SC0025809. 
F.E.-O. was supported by the US National Science Foundation under grant No. PHY-2310656. C.L. is exclusively supported by MERL. We thank Costin Iancu and Ed Younis for assisting with the BQSkit optimization. This research was conducted using Pinnacles (NSF MRI, No. 2019144) at the Cyberinfrastructure and Research Technologies (CIRT) at University of California, Merced.  We acknowledge the use of IBM Quantum Credits for this work. The views expressed are those of the authors, and do not reflect the official policy or position of IBM or the IBM Quantum team. 
}

\appendix
\renewcommand{\thesection}{Appendix~\Alph{section}}

\section{Dynamics simulation via Trotterization on classical hardware}\label{App:Dt}
To demonstrate that the dynamics obtained by Trotterization with $\Delta t/t_{0}=0.1$ closely approximates higher-order methods, we solve the Schrodinger equation for a five-spin TFIM with an initial state of $\ket{\psi_0}=\ket{0}^{\otimes 6}$  up to $\tau/t_{0}=25$ using second-order Trotterization shown in Eq.~\eqref{eq:second_trotter_tfim} on a classical computer. The results computed by Trotterization are compared with the fourth-order Runge-Kutta algorithm with $\Delta t/t_{0}=10^{-3}$ (black-line) on a classical computer by plotting the population $|\langle 0^{\otimes 6}|\psi(t)\rangle|^{2}$.

\begin{figure}[t]
    \centering
    \includegraphics[width=0.7\linewidth]{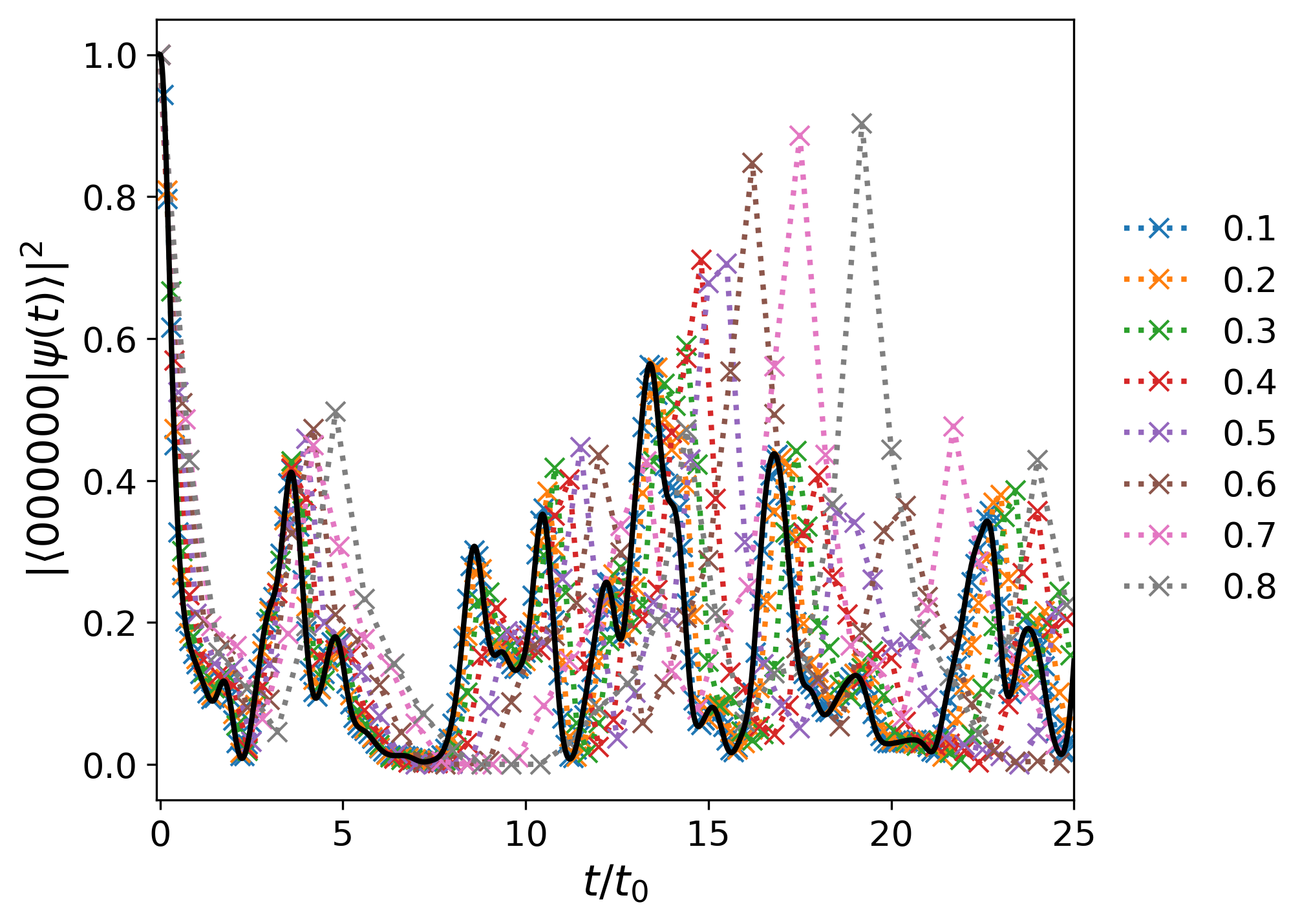}
    \caption{The population of $|\langle 0^{\otimes 6}|\psi(t)\rangle|^{2}$ for a six-spin TFIM at various $\Delta t/t_{0}$ by classical simulation using second-order Trotterization. The initial state is $\ket{\psi_0}=\ket{0}^{\otimes 6}$. The black line is the fourth-order Runge-Kutta with $\Delta t/t_{0}=10^{-3}$. The evolution for $\Delta t=0.1t_{0}$ shows little deviation from the black line.}
    \label{fig:TrotterizationTFIM_N6}
\end{figure}

Fig.~\ref{fig:TrotterizationTFIM_N6} shows that for long time evolution the second-order Trotterization deviates from the exact curve at values of $\Delta t/t_{0}>0.1$. Only $\Delta t/t_{0}=0.1$ remains tightly bound to the numerically exact solution (black-line). Furthermore, choosing a large step-size $\Delta t/t_{0}$ shifts the peaks in the curves. Since our estimation method relies on observing the periodicity of theses peaks via Fourier transform, broadening of the period will shift our peaks in the frequency domain and introduce additional error in our energy estimation. Therefore, we have chosen $\Delta t/ t_{0}=0.1$ as our time-step for the TFIM.

\begin{figure}[t]
    \centering
    \includegraphics[width=0.7\linewidth]{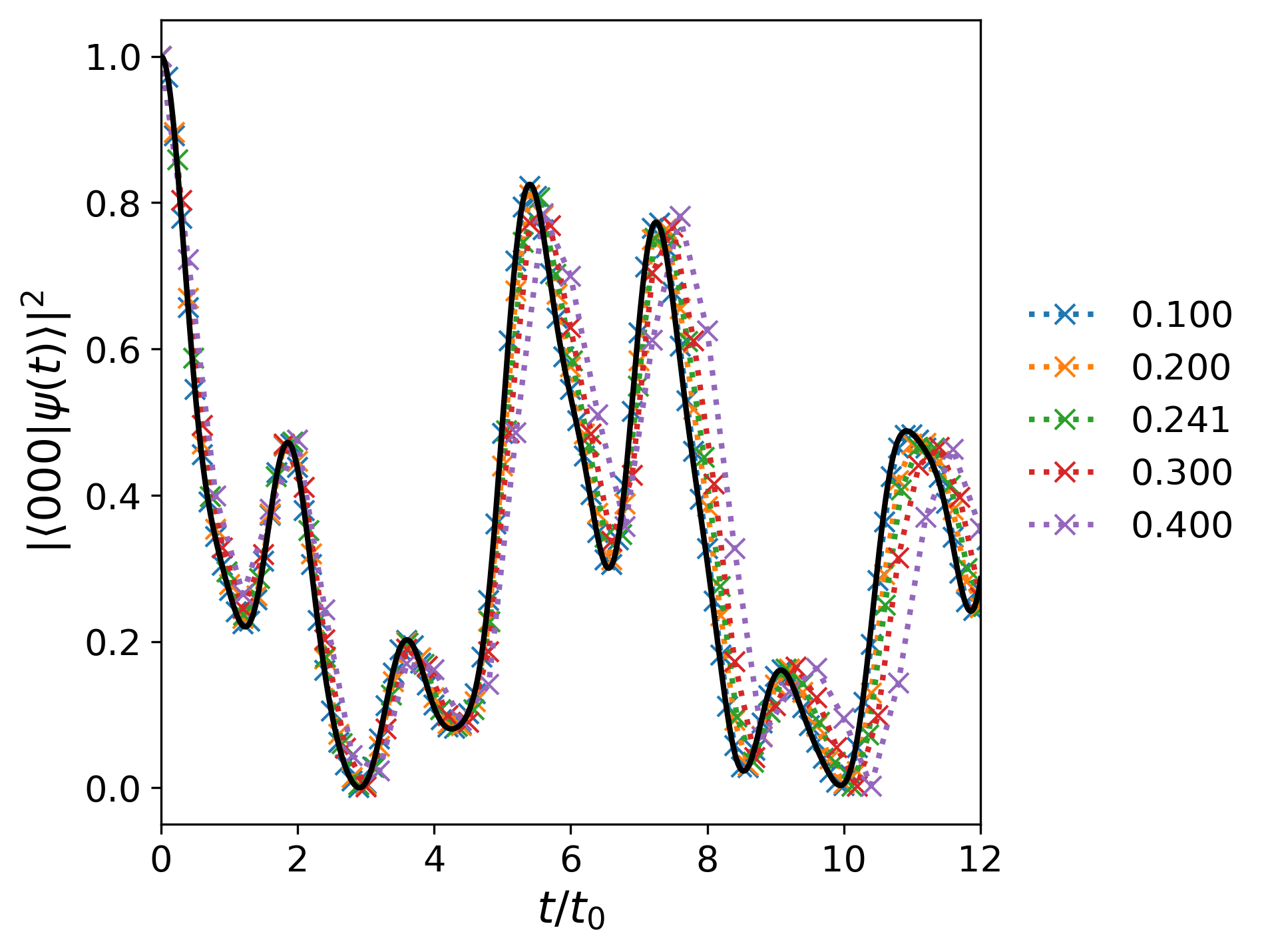}
    \caption{The population of $|\langle 0^{\otimes 3}|\psi(t)\rangle|^{2}$) for a three-spin TLFIM at various $\Delta t/t_{0}$ by classical simulation using second-order Trotterization. The initial state is $\ket{\psi_0}=\ket{0}^{\otimes 3}$. The black-line is the fourth-order Runge-Kutta with $\Delta t/t_{0}=10^{-3}$. The evolution for $\Delta t=0.241t_{0}$ shows minimal deviation from the black line.}
    \label{fig:TrotterizationTLFIM_N3}
\end{figure}

In contrast to the TFIM, a larger step size of ($\Delta t/t_{0}=0.241$) was chosen for the three-spin TLFIM to mitigate the increasing circuit depth associated with the implementation of the second-order Trotterization. Additionally, this choice ensures that the ground-state energy is commensurate with the frequency spacing and maintains the computational cost of BQSKit optimization tractable.
To assess the accuracy of this coarse discretization, we again compare the Trotterized dynamics in Eq.~\eqref{eq:second_trotter_tfim} with the fourth-order Runge-Kutta methods on classical computers. The three-spin TLFIM is initialized in the state $\ket{\psi_0}=\ket{0}^{\otimes 3}$ and evolved up to $\tau/t_{0}=M\Delta t\approx12$ ($M=50$). The comparison is performed by plotting the population  $|\langle 0^{\otimes 3}|\psi(t)\rangle|^{2} $. As shown in Fig.~\ref{fig:TrotterizationTLFIM_N3}, the Trotterized dynamics obtained with $\Delta t/t_{0}=0.241$ exhibit very minor deviations from the Runge-Kutta solution on classical computers. 
The choice of $\Delta t/t_{0}=0.241$ was originally motivated by practical considerations, where larger time steps helped limit circuit-depth and optimization costs. Following the discovery of a CDC for the three-spin TLFIM, $\Delta t/t_{0}$ can in principle be reduced to allow for more accurate long-time evolution with improved spectral resolution.

\section{Matchgate Structure}\label{App:Matchgate}
The matchgates form a group of two-qubit gates $G(A, B)$ with the following form
\begin{equation}
    G(A, B) = \begin{pmatrix}
        p & 0& 0& q\\
        0 & w& x& 0\\
        0 & y& z& 0\\
        r & 0& 0& s
    \end{pmatrix},  
    A = \begin{pmatrix}
       p& q\\ 
       r& s
    \end{pmatrix},    
    B = \begin{pmatrix}
       w& x\\ 
       y& z
    \end{pmatrix},
\end{equation}
where $A$ and $B$ are in $SU(2)$ or $U(2)$ with the same determinant. $A$ acts in the even parity subspace $span\{\ket{00}, \ket{11}\}$ and $B$ in the odd parity subspace $span\{\ket{01}, \ket{10}\}$. In particular, the Hilbert space decomposes into the direct sum of the even and odd parity subspaces, on which the operator acts as $A \oplus B$~\cite{Jozsa2008}. Furthermore, the group structure allows for adjacent gates to be merged:
\begin{center}
\begin{quantikz}[row sep=0.1cm]
&\gate[2]{G}& \gate[2]{G}&\\
& & &\\
\end{quantikz}
=
\begin{quantikz}[row sep=0.1cm]
&\gate[2]{G}&\\
& &\\
\end{quantikz}.
\end{center}
Furthermore, the mirroring property has been conjectured by numerically optimizing the parameters on both sides of the circuit~\cite{BassmanOftelie2022}
\begin{center}
\begin{quantikz}[row sep=0.1cm]
&\gate[2]{G}& &\gate[2]{G}&\\
& &\gate[2]{G}&&\\
& & & &\\
\end{quantikz}
=
\begin{quantikz}[row sep=0.1cm]
& &\gate[2]{G}& & \\
&\gate[2]{G} & &\gate[2]{G} & \\
& & & &\\
\end{quantikz}
\end{center}
The mirroring and group properties allow for the $M$ trotter step quantum circuit to be down folded into a $N$-qubit quantum circuit with $N$ layers of matchgates, where the number of layers $N<M$ for longer time evolution. Circuits with an odd number of qubits have $N\lfloor
\frac{N}{2}\rfloor$ number of matchgates, and even number of qubits contain $\frac{N}{2}(N-1)$ matchgates. However, this is limited to a subset of spin Hamiltonians that are quadratic in fermionic operators~\cite{BassmanOftelie2022}.

\section{Comparison of quantum dynamics from classical and quantum computers}
As seen in Table.~\ref{tab:tfim_isa_gate_count} and $N=3$ in Table.~\ref{tab:tlfim_isa_gate_count}, the total ISA gate count for the unitary gate remains well below the NISQ limit. The $N=7$ TFIM has the largest gate count, at 217 gates. Since the unitary gate corresponds to the time-evolution operator, these results suggest that the dynamics of the $N=3-7$ TFIM and the $N=3$ TLFIM can be reliably simulated on current quantum hardware, with minimal noise.

\begin{figure}[t]
    \centering
    \includegraphics[width=0.7\linewidth]{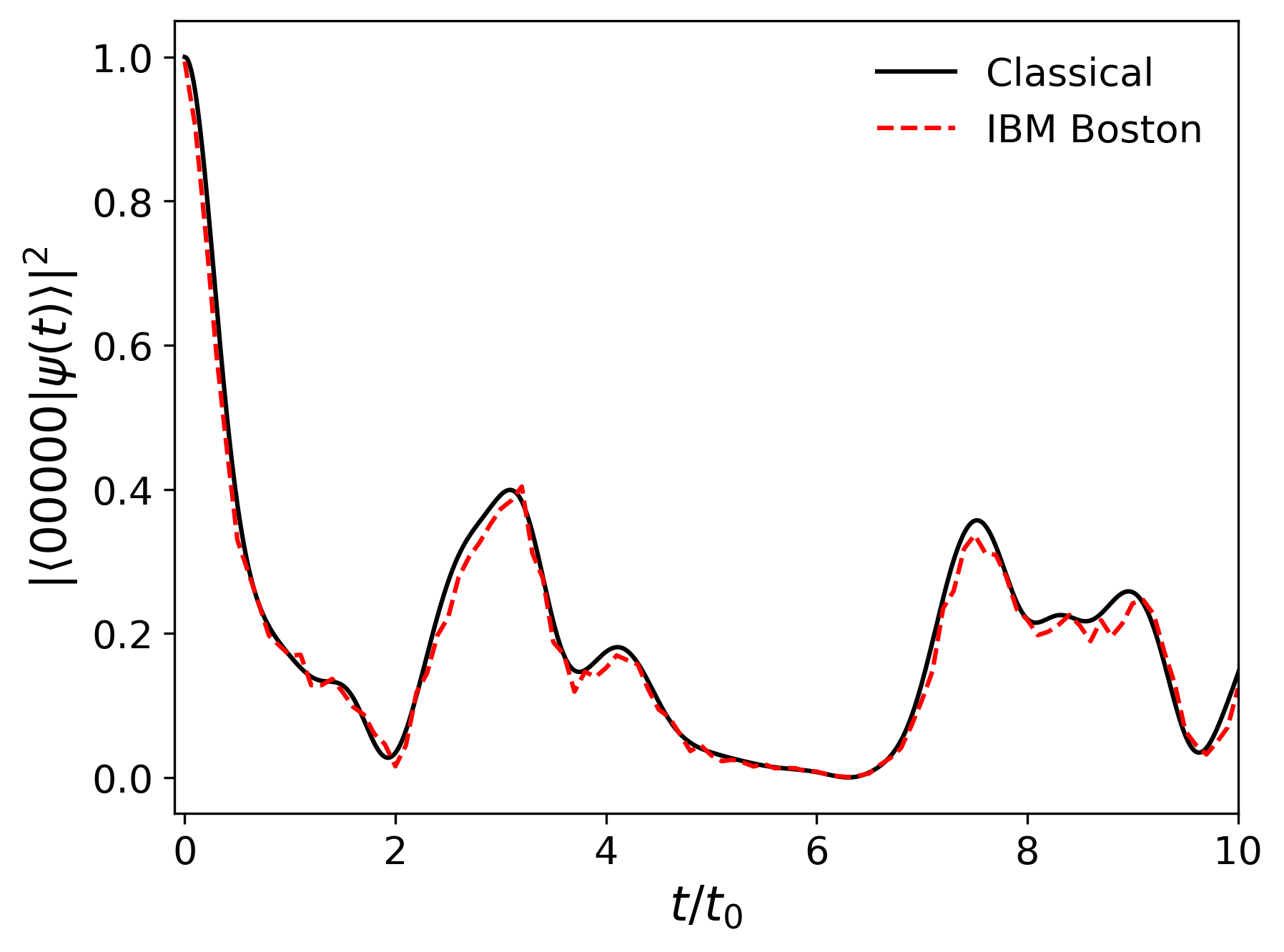}
    \caption{The population of $|\langle 0^{\otimes 5}|\psi(t)\rangle|^{2}$) for a five-spin TFIM with $\Delta t/t_{0}=0.1$ and $\tau/t_{0}=10$. The red-dotted line is the population obtained from IBM Boston, which is compared to the classical simulation using fourth-order Runge-Kutta with $\Delta t/t_{0}=1e^{-3}$. The initial state is $\ket{\psi_0}=\ket{0}^{\otimes 5}$, and a total of 101 CDCs were ran with 6500 shots per circuit.}
    \label{fig:tfim_dyn}
\end{figure}

\begin{figure}[t]
    \centering
    \includegraphics[width=0.7\linewidth]{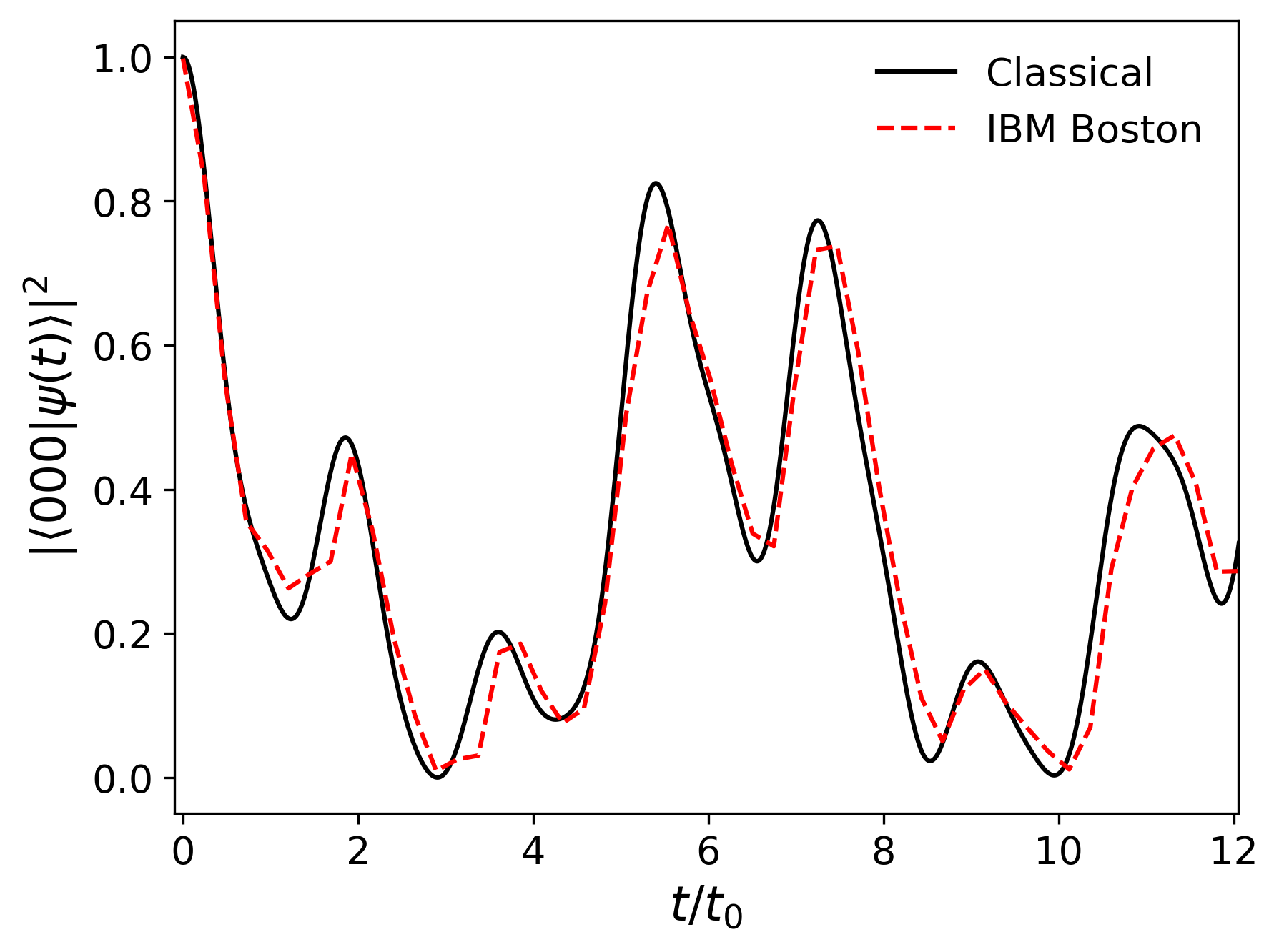}
    \caption{The population of $|\langle 0^{\otimes 3}|\psi(t)\rangle|^{2}$) for a five-spin TLFIM with $\Delta t/t_{0}=0.241$ and $M=50$ Trotter steps ($\tau/t_{0}\approx 12$). The red-dotted line is the population obtained from IBM Boston, which is compared to the classical simulation using fourth-order Runge-Kutta with $\Delta t/t_{0}=10^{-3}$. The initial state is $\ket{\psi_0}=\ket{0}^{\otimes 3}$, and a total of 51 CDCs were ran with 6500 shots per circuit.}
    \label{fig:tlfim_dyn}
\end{figure}

We verify the resolution of the dynamics for a $N=5$ TFIM and $N=3$ TLFIM. For the $N=5$ TFIM, we run the CDCs obtained from optimization with the set of unitaries from second-order Trotterization (See Section II C). The TFIM is evolved to a final time of $\tau/t_{0}=10$ with $\Delta t/t_{0}=0.1$, which corresponds to $M=100$ CDC circuits. We measure the population $|\langle 0^{\otimes 5}|\psi(t)\rangle|^{2}$ on IBM Boston, and compare against the fourth-order Runge-Kutta method with a fine time-step $\Delta t/t_{0}=10^{-3}$. Although the TFIM is evolved to $\tau/t_{0}=50$ in the main text, we evolve the system to shorter time due to the limited QPU time.
Likewise, we obtain the dynamics for the $N=3$ TLFIM. For the TLFIM, we execute the circuits from global BQSkit optimization (Section III B). The system contains $M=50$ Trotter steps with a time-step of $\Delta t/t_{0}=0.241$, which corresponds to a final evolution time near $\tau/t_{0}\approx 12$. Again, this is compared to the population of $|\langle 0^{\otimes 3}|\psi(t)\rangle|^{2}$ computed by the fourth-order Runge-Kutta method with a fine time-step $\Delta t/t_{0}=10^{-3}$ on classical computers.
As seen in Fig.~\ref{fig:tfim_dyn} and  Fig.~\ref{fig:tlfim_dyn}, the population $|\langle 0^{\otimes N}|\psi(t)\rangle|^{2}$ from IBM Boston follows the dynamics from the fourth-order Runge-Kutta method on classical computers very well with little deviation. 

\section{Fermionic representation of the TLFIM}\label{App:fermion_TLFIM}
Here we show that the Hamiltonian of the TLFIM does not map to quadratic fermions, a requirement for CDCs. The Hamiltonian of the TLFIM is shown in Eq.~\eqref{eq:tlfim}.
The spins are mapped onto fermions using the standard Jordan-Wigner transformation~\cite{Batusta2001JW}.
\begin{gather}
    \sigma^{j}_{z} = (c^{\dagger}_{j} +c_{j})\prod_{k=1}^{j-1}(1-2c^{\dagger}_{k}c_{k}),\\
    \sigma_{x}^{j} = \mathbb{I} - 2c_{j}^{\dagger}c_{j}.
    \label{eq:jordan-wigner}
\end{gather}
Here $c_{j}^{\dagger}$ ($c_{j}$) is the fermion creation (annihilation) operator at site $j$.
It is well known that the Hamiltonian $\hat{H}$ of the TFIM shown in Eq.~\eqref{eq:tfim} is quadratic under Jordan-Wigner transformation~\cite{Pierre19701DTfim}. Explicitly,
\begin{gather*}
    \hat{H} =-h_{x}(\mathbb{I} - 2c_{N}^{\dagger}c_{N})-J_{z}\sum_{j}^{N-1}(c^{\dagger}_{j} - c_{j} )(c^{\dagger}_{j+1} +c_{j+1})\\
    -h_{x}\sum_{j}^{N^-1}(\mathbb{I} - 2c_{j}^{\dagger}c_{j}).
\end{gather*}
For the TLFIM described in Eq.~\eqref{eq:tlfim}, it is clear that we need to analyze the effect from the extra z-field.  Under the Jordan-Wigner transformation, it can be shown that the z-field term has cubic fermion operators
\begin{equation}
    \sum_j^N\sigma_z^j =  \sum_j^N(c^{\dagger}_{j} +c_{j})\prod_{k=1}^{j-1}(1-2c^{\dagger}_{k}c_{k}).
\end{equation}
 Thus, the Hamiltonian of the TLFIM is not quadratic in fermionic representation, and therefore, its time-evolution operator $e^{-i\hat{H}_{TLFIM}t}$ may consist of non-n.n gates, i.e., non-matchgates. Nevertheless, we found the CDCs for $N=3$ TLFIM through BQSkit optimization.

\clearpage

\bibliographystyle{unsrt}

\end{document}